\documentclass[a4paper,12pt,reqno,superscriptaddress,nofootinbib]{revtex4}
\usepackage[centertags]{amsmath}
\usepackage{amsfonts}
\usepackage{amssymb}
\usepackage{amsthm}
\usepackage{newlfont}
\usepackage{stmaryrd}
\usepackage{mathrsfs}
\usepackage{euscript}
\usepackage{siunitx}
\usepackage{comment}
\usepackage{algorithm2e}
\usepackage{graphicx,subcaption}
\usepackage{enumitem}
\usepackage{natbib} 
\usepackage{braket}
\usepackage{bbm}
\usepackage{graphicx}
\usepackage{xcolor}
\usepackage{floatrow}
\usepackage{caption}
\usepackage{hyperref}
\usepackage{hhline}
\usepackage{hyperref}
\usepackage{cleveref}
\usepackage{amsmath}
\usepackage[bb=dsserif]{mathalpha}
\usepackage{bm}

\theoremstyle{plain}
  \newtheorem{theorem}{Theorem}[section]

\theoremstyle{definition}

\theoremstyle{remark}

  \newtheorem{example}[theorem]{Example}

\numberwithin{equation}{section}

\let\ve=\varepsilon

\newcommand{\opunit}{\text{1}\kern-0.22em\text{l}}

\DeclareMathAlphabet{\mathpzc}{OT1}{pzc}{m}{it}

\newcommand{\fig}{Fig.\;}

\newcommand{\id}{\textrm{d}}

\usepackage{floatrow}
\usepackage{caption}
\DeclareCaptionJustification{justified}{\leftskip=0pt \rightskip=0pt \parfillskip=0pt plus 1fil}

\numberwithin{equation}{section}

\usepackage{xcolor,soul}

\newcommand{\tg}{\textcolor{teal}}

\begin{document}
\title{Negative specific heats:\\ where Clausius and Boltzmann entropies separate}
\author{Lander Bogers, Faezeh Khodabandehlou, and Christian Maes\\
{\it Department of Physics and Astronomy, KU Leuven}}
\begin{abstract}
Steady nonequilibria dissipate energy and, when changing external parameters, an extra or excess heat accompanies the relaxation to the new nonequilibrium condition.  For nonequilibrium systems in contact with a thermal bath, the heat capacity is defined as that excess heat per degree temperature for a quasistatic change of the bath temperature.
It is fairly common to find negative heat capacities for steady nonequilibrium systems, in contrast with the situation for systems in thermal equilibrium. We discuss and illustrate the origin of that negative thermal response using Markov models.  We find that the negativity results from an anticorrelation between quasipotential and (a change in) pseudopotential, the first measuring (excess) heat (and Clausius entropy), and the latter being related to the Boltzmann entropy. It can be quantified via an appropriate choice of effective temperatures.
    \end{abstract}

\maketitle
\section{Introduction}
There is only one entropy for macroscopic equilibrium systems. The original understanding of the Clausius or thermodynamic entropy $S$ derives from the well-known formula $\delta Q^\text{rev}/T = \id S$, part of the famous heat theorem from which reversible heat over temperature is an exact differential.  While the kinetic interpretation of temperature marked the beginning of statistical mechanics, the notion of heat got its role and place in bookkeeping changes in energy (First Law of Thermodynamics).  So many years later, the work of Boltzmann, Planck and Einstein gave a totally new and statistical meaning to entropy, where we now write $S = k_B \log W$ for the Boltzmann entropy counting the microscopic degeneracy $W$ of a macroscopic condition (at some fixed energy etc).  Miraculously, Boltzmann and Clausius entropies (and others like the Gibbs ensemble entropy) agree under thermodynamic equilibrium, at least for quasilocal interactions. It implies that fluctuations, governed by the Boltzmann entropy, and dissipation, represented by the Clausius entropy, are closely linked as also made explicit in response relations following from the fluctuation-dissipation theorem.  Yet, we cannot expect that to remain true and unchanged when we move far away from equilibrium, \cite{joel,david,jona}.  For steady nonequilibrium conditions, it is, first of all, unclear how to define the analogues of temperature or entropy, and, more generally, the standard fluctuation-dissipation relations are simply violated when well away from equilibrium, \cite{scr,proS}.  In other words, Boltzmann and Clausius entropies are bound to separate. The main point of the present paper is to make that statement operational via the study of nonequilibrium heat capacities.  Their negativity is the sign of separation, as we will see that the negativity of the heat capacity indicates an anticorrelation between the excess heat and the change in population degeneracy.\\

The thermal response refers to the behavior of a physical system under changes in temperature or when subjected to heat pulses.  For a macroscopic system in equilibrium, one can measure the change in its temperature by heating it.  That can be done under various constraints $Z$ such as, {\it e.g.} for a gas keeping its volume or pressure constant, and gives rise to the standard definitions of (equilibrium) heat capacities $C_Z(T)$.  In short, $\delta Q_Z^\text{rev} = C_Z(T)\, \id T$ where $Q_Z^\text{rev}$ is the reversible heat given to the system to increase its temperature by $T\rightarrow T+\id T$ while keeping the constraint $Z$.  Interestingly, heat capacity may on the one hand be used to measure energy, entropy, or enthalpy, and on the other hand inform about the variance (fluctuations) of those potentials in the corresponding equilibrium ensemble.  However, those relations are not given when the interaction becomes long-ranged such as for Newtonian gravity or when the system is nonthermodynamic such as in  finite-size clusters,  \cite{LYNDENBELL1999293,PhysRevLett.86.1191,PhysRevLett.91.130601}, where the equivalence of ensembles is violated; negative heats become possible there.\\
The situation for nonequilibrium heat capacities is conceptually similar but results in other expressions.  We refer to \cite{simon, jchemphys, closeheat,cejp,epl,calo,TLSpre2024} for introductions and examples.\\

Section \ref{secheatc} reminds the reader of the relevant setup.  For nonequilibrium steady conditions, we need an open system, and we identify a thermal bath in its environment where the system dissipates heat. Upon small and slow changes $\id T$ in the bath, an excess of heat $\delta Q^\text{exc} = C(T)\id T$ is absorbed by the system that defines the heat capacity $C(T)$ at temperature $T$ of the thermal bath.  We ignore from now on in the notation the possibility of different constraints $Z$, which can act on the system itself and on the bath.
It is no longer true that $C(T)$ needs to be positive and indeed we know plenty of examples where $C(T) <0$.  It is an interesting feature which goes hand in hand with the typical extra we get from nonequilibrium heat capacities: $C(T)$ is able to pick up dynamical or kinetic information about the system which is not available when scanning it in equilibrium.\\ 
Section \ref{frr} gives the theoretical framework for understanding the occurrence and the implication of negative specific heats for nonequilibrium systems. The heat capacity gets written as a covariance in the stationary distribution.  The two involved random variables are anticorrelated when the heat capacity is negative.  They are related on the one hand to the quasipotential or excess heat, which gives the energy in the case of equilibrium, and, on the other hand, to the pseudopotential, which gives the population statistics.  It is there that we see how Clausius and Boltzmann entropies separate.\\
Section \ref{effective} defines the relevant effective temperatures and we use it to quantify the negativity of the heat capacity.  In particular, a sufficient relation follows for that negativity.\\
Finally, in Section \ref{illu} we present a number of discrete models with agitated and double-channel transitions to illustrate the theory and the specific origin of negativity at very low and at intermediate temperatures. Each time, we see that the probability of the quasi-ground state (minimizing the quasipotential or expected excess heat) is increasing with temperature (quite unlike the situation for systems in thermal equilibrium) and the effective temperatures (defined with respect to that quasi-ground state) are decreasing with temperature.  It indicates a population anomaly (even leading to an inversion): as the temperature of the bath increases, low-lying quasi-energies get more populated.  It can also happen without population anomaly when the effective temperatures become almost constant as function of the bath temperature (again, unlike in equilibrium).
 
\section{Nonequilibrium heat capacity}\label{secheatc}

\begin{figure}[H]
\centering
\includegraphics[scale=0.5]{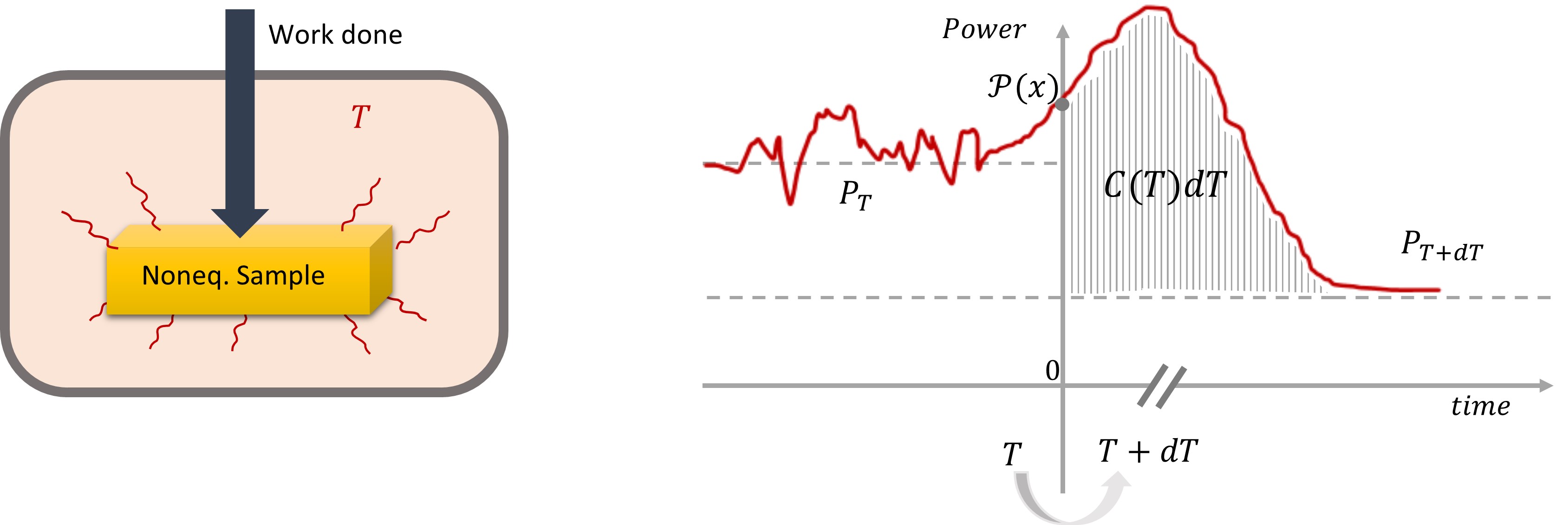}
\caption{\small{Cartoon of setup.  Left: irreversible work is done on a dissipative sample, bringing it in a stationary nonequilibrium condition.  It dissipates into a thermal bath at temperature $T$.  Right: Before time zero the dissipated power is variable and fluctuates around its steady value ${\cal P}_T$.  At time zero, the nonequilibrium system is at state $x$ and after slowly perturbing the temperature of the bath, the system relaxes to a new steady condition with a new value for the steady dissipated power. The shaded area gives the excess heat which is a fluctuating quantity (with $x$) and the average per $\id T$ is the heat capacity.}} \label{setup}
\end{figure}

Suppose that we control a steady nonequilibrium sample in contact with a heat bath at temperature $T$; see Fig.\ref{setup}.  The sample is an open system on which irreversible work is being done, either via its boundaries imposing conflicting conditions, or in the bulk via nonconservative or time-dependent forces, or via internal fueling of particles as in active matter.  Whatever the case, there will be a steady dissipated power of (Joule) heat flowing into the bath, as suggested in the left cartoon of Fig.\ref{setup}.  Clearly, the heat flow may fluctuate in time, but on average it is positive and depends on many parameters such as the (fixed) temperature $T$ of the bath.  Heat capacity is about how that flux depends on $T$.  More specifically, we need the excess heat in a quasistatic temperature variation, \cite{oono}. The shaded area in the right of Fig.\ref{setup} is a measure of that heat capacity. 
 For more precision and mathematical details, we refer to the original papers \cite{epl,cejp} and to more recent publications \cite{simon, jchemphys, closeheat, calo,TLSpre2024}.  Experiments are in progress, with \cite{cera,cerro89} as early measurements.\\

The theoretical modeling so far is restricted to Markov processes.  We have here time-homogeneous processes $X_t$ that converge exponentially fast to a unique stationary distribution $\rho(x), x \in K$ over state space $K$, satisfying an equation of the form $L^\dagger \rho =0$ where $L^\dagger$ is the forward generator.  The backward generator $L$ generates the time evolution in the sense that
\[
e^{tL}g(x) = \langle g(X_t)\,|\,X_0=x\rangle
\]
for state functions $g$. The average $\langle\cdot\rangle$ is a process expectation. That abstract setting gets meaning within thermal physics from the moment when the notion of ``expected dissipated heat in a given state'' makes physical sense. That is connected with the notion of local detailed balance, \cite{ldb,time}.  It includes the identification of a heat bath at temperature $T$; the whole process is now parametrized by $T$.  Let us denote the expected heat by  ${\cal P}_T(x) = \cal P(x)$ for state $x$, and write $P_T = P = \langle \cal P(x)\rangle_T $ for its stationary value (with stationary distribution $\rho_T$ and stationary expectations denoted by $\langle \cdot\rangle_T$).  Define the quasipotential $V_T=V$ as the centered function, $\langle V_T\rangle_T =0$, that solves the Poisson equation, \cite{pois},
\begin{equation}\label{poisson}
    LV(x) +  \cal P(x) - P =0, \quad V(x) = \int_0^\infty \id t\, e^{tL}(\cal P(x)-P)
\end{equation}

The heat capacity $C(T)$ is then obtained as
\begin{equation}\label{ct}
C(T) =  - \Big\langle \frac{\partial}{\partial T} V_T \Big\rangle_T,\quad \text{or}\quad \sum_{x\in K} \rho(x) \frac{\partial}{\partial T} V(x) 
\end{equation}
where the last formula holds for a finite state space $K$.  Note that the stationary distribution $\rho$ is by and large not explicitly known.  However, under global detailed balance (equilibrium dynamics), say at a fixed volume, there is an energy state function $E$ for which 
$\cal P(x) = LE(x)$, and thence, $V(x) = E(x) - \langle E\rangle_T$, which reduces \eqref{ct} to the well-known equilibrium formula for the heat capacity at fixed volume.\\

The interest in the notion of nonequilibrium heat capacity derives from the wish to quantify thermal response, and to understand what information is encoded in it concerning the nonequilibrium condition and its dynamics.  A variety of toy-examples have been studied so far, \cite{simon,epl,cejp,pritha,TLSpre2024}, and sufficient conditions have been formulated for a nonequilibrium extension of the Third Law of thermodynamics, \cite{jchemphys}.  It has also been observed that this nonequilibrium heat capacity can become negative, which makes the question of the present paper: to understand the meaning and the origin of that negativity.  We focus on models with a discrete state space, that show a typical Schottky anomaly already for zero driving, \cite{schottky,Tar2003TheSH}.  Some driven examples of Section \ref{illu} show an inverted Schottky anomaly, visible already in the right plot of Fig.~\ref{hc3l} for the next example.
For examples of Markov diffusions with negative specific heats, we refer to \cite{pritha, epl, cejp}; the understanding of negativity there is the same as for the jump processes that we consider below.\\

\begin{example}\label{ex1}
    Consider an agitated molecular system, where the hierarchy of energy levels randomly switches. Such a molecular switch can be modelled as a Markov jump process on a ladder with (to be specific) three levels; see \fig \ref{barrier}. Each leg $\sigma=\pm $ has three states $\eta = 1,2,3$, so that the states are of the form $x=(\eta,\sigma)$. Each state located in $\sigma=-$ has an energy $E(\eta,-) =(\eta-1)\ve $  and the states on $\sigma=+$ has an energy  $ E(\eta, +)=(3-\eta)\ve $, and the process is switching legs at rate $\alpha$. The transition rates are chosen as
    \begin{align}\label{rate1}
  & k((1,-),(2,-))=k((2,-),(3,-))=k((3,+),(2,+))=k((2,+),(1,+))= e^{-\beta(\Delta + \ve/2)}\notag \\
  & k((1,+),(2,+))=k((2,+),(3,+))=k((3,-),(2,-))=k((2,-),(1,-))= e^{-\beta(\Delta - \ve/2)} \notag\\
 & k((\eta,-),(\eta,+))=k((\eta,+),(\eta,-))=\alpha 
\end{align}
where $\Delta>0$ is an energy barrier,  $(\eta,\sigma)$ denotes a state on the leg $\sigma$ and level $\eta$ and  $k((\eta ,\sigma),(\eta',\sigma))$ denotes the transition rate from state $(\eta ,\sigma)$ to state $(\eta',\sigma)$.\\ 

\begin{figure}[H]
\centering
\includegraphics[scale=0.65]{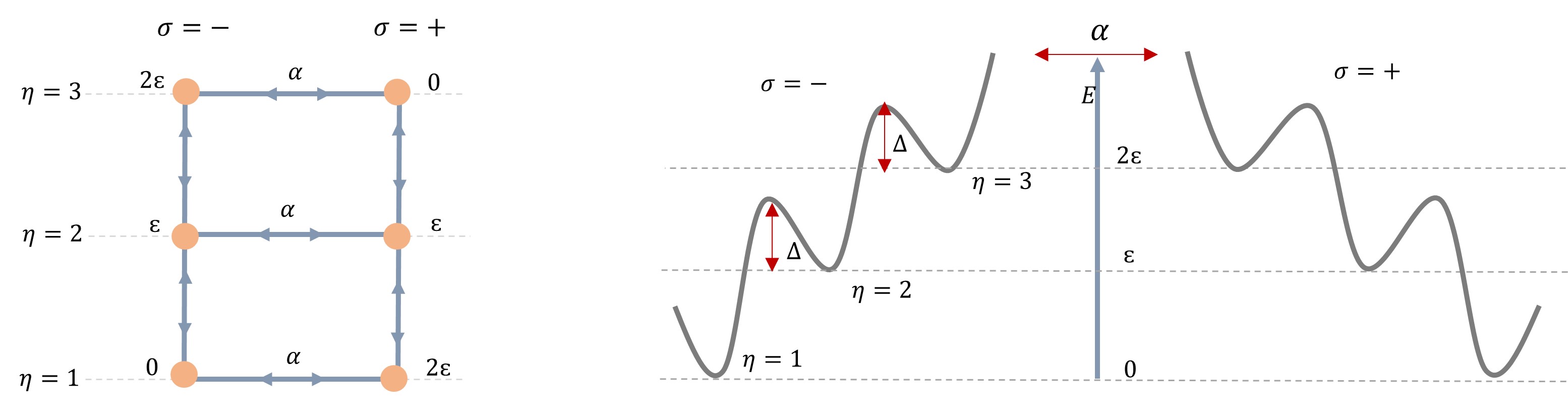}    
\caption{\small{Left:  three-level ladder.  Right: the energy landscape, where $\Delta$ denotes the height of the barrier. A molecular switch can indeed also be reproduced experimentally with a colloid moving in an optically simulated and flashing landscape.}}   \label{barrier}
\end{figure}
The heat flowing to the bath at inverse temperature $\beta$ equals $\ve$ at each transition $\eta\rightarrow \eta'$ where the level is changing. The changing of legs is  work done by external sources.\\
For completeness, the stationary distribution, the quasipotential and the heat capacity of this three-ladder are calculated in Appendix \ref{3level}. The heat capacity is plotted in \fig\ref{hc3l}. Taking $\alpha=0$  corresponds to the equilibrium case.  For  $\alpha >0 $, the nonequilibrium heat capacity depends kinetically on the barrier $\Delta$ and may become negative (here, at low temperatures for large enough $\Delta$).
\begin{figure}[H]
     \centering
      \begin{subfigure}{0.49\textwidth}
         \centering
         \def\svgwidth{0.8\linewidth}        
        \includegraphics[scale = 0.85]{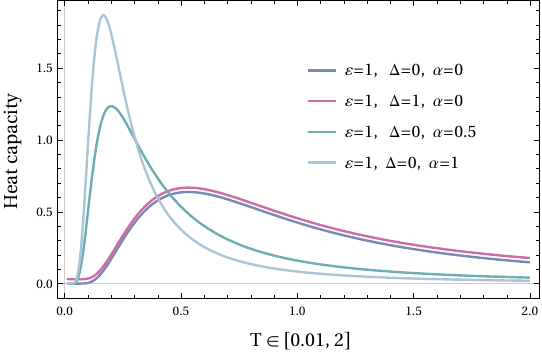}
     \end{subfigure}
     \hfill
     \begin{subfigure}{0.49\textwidth}
         \centering
         \def\svgwidth{0.8\linewidth}        
   \includegraphics[scale = 0.85]{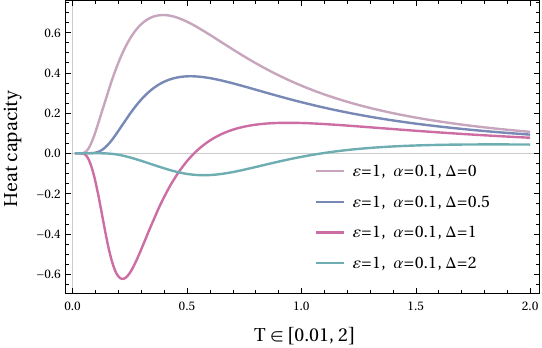}
     \end{subfigure}
\caption{\small{Heat capacity of a $3$-level ladder  as a function of temperature $T$ for different values of  $\alpha, \Delta $ and $\ve$ as defined in the transition rates \eqref{rate1}. We observe an inverted Schottky anomaly for large enough $\Delta$ when $\ve=1$.}}   \label{hc3l}
\end{figure}
\end{example}
This example is continued in Section \ref{cont}.

\section{Heat capacity as covariance}\label{frr}
For Markov jump processes with a finite state space $K$, we can rewrite the heat capacity \eqref{ct} as
\begin{equation}\label{hca}
C(T)\,\id T =  - \id T\,\Big\langle \frac{\partial}{\partial T} V_T \Big\rangle_T = 
\sum_{x\in K} V_T(x) \,\rho_{T+\id T}(x) = \lim_{t\uparrow \infty} \big\langle V_T(X_t)\big\rangle_{T+\id T} 
\end{equation}
where we have used that the quasipotential $V_T$ is centered, $\sum_x V_T(x) \,\rho_{T}(x) =0$, and where the last expectation $\langle\cdot\rangle_{T+\id T}$ is for the dynamics at temperature $T+\id T$. Under detailed balance, the Kubo formula leads from \eqref{hca} to the variance of the energy since then and there, $V_{T} = E - \langle E\rangle_T$.  Nonequilibrium response theory is more complicated as it does not only involve a correlation with the heat but also with the frenesy.
The resulting expression is formally complicated and not revealing much specific detail about the negativity of the heat capacity, except that it is clearly the frenetic contribution in general that must be responsible for the negativity.\\

Another road to express the heat capacity as a correlation function is less explicit but conceptually more attractive.
In fact, the heat capacity \eqref{ct} or \eqref{hca} can be written as 
\begin{equation}\label{hcaf}
C(T) =  \bigg\langle \frac{\id \log\rho_T }{\id T}\, V_T  \bigg\rangle_T 
\end{equation}
That is a stationary correlation function between, on the one hand,$V_T(x) = \delta Q^\text{exc}(x)$ which  is directly related to an expected excess heat when in state $x$, and on the other hand $\frac{\id  }{\id T}\log\rho_T(x)$ which is the change in state probability with temperature. We can even make it a covariance between a heat capacity $C_B(x)$ derived from the Boltzmann entropy, and a Clausius entropy $S_C(x) =  \delta Q^\text{exc}(x)/T$,
\begin{equation}
C(T) = \bigg\langle C_B\,;\,S_C  \bigg\rangle_T,\qquad \text{ for }  C_B(x) = T\frac{\id \log\rho_T(x) }{\id T}
\end{equation}
We therefore introduce the pseudopotential $\Phi(x) = \log \rho(x)$, mimicking the Boltzmann entropy, and the heat capacity \eqref{hcaf} gets rewritten as
\begin{equation}\label{hcal}
C(T) =  \bigg\langle \frac{\id \Phi_T }{\id T}\, V_T  \bigg\rangle_T 
\end{equation}
 We know that, for equilibrium systems, increasing the bath temperature will increase the absorbed heat of the system and will thereby cause an increased equidistribution of the finite number of energy levels.  In particular, we expect the lowest energy state to lose occupation.  That connects heat with the density of states, and enables, always around thermal equilibrium, to measure entropy via montoring the exchanged heat. Heat is leaving the system if higher energy states become less occupied and {\it vice versa}. This is what makes the equilibrium heat capacity positive: mathematically and corresponding to \eqref{hcal}, in equilibrium (and in fact also close to equilibrium, \cite{closeheat}), we have $\id \Phi_T /\id T =  V_T/(k_BT^2)$.\\
Inspecting \eqref{hcal}, the negativity of the nonequilibrium heat capacity signifies at least an anticorrelation between the heat absorbed and the change in occupation with increasing bath temperature.  One possibility is that increasing the temperature may, at least in certain regimes, lead to higher occupation of low-energy states. That is not unrelated to the idea of negative temperatures for gravitating or for very inhomogeneous and smaller systems for which the potential energy is anticorrelated with the kinetic energy and/or that fail to satisfy equivalence of ensembles, \cite{LYNDENBELL1999293,PhysRevLett.86.1191,PhysRevLett.91.130601}. 
What happens is that increasing the temperature may, at least in certain regimes, lead to higher occupation of low-energy states.

\section{Negative heat capacity via effective temperatures}\label{effective}

Although the covariance identity \eqref{hcaf} (or \eqref{hcal}) helps to conceptualize the negativity of the heat capacity, here we look for a sufficient and quantitative condition in terms of effective temperatures.  We consider the case where there is a well-defined energy $E(x)$ associated to each state, so that in equilibrium
the occupation
\begin{equation}\label{bg}
\rho_\text{eq}(x) \sim e^{-\beta E(x)}
\end{equation}
follows the Boltzmann-Gibbs statistics.\\

Fixing the bath temperature $T$, we denote by $x^*$ any specific state that has the smallest quasipotential: for all $x$,
\[
 \Delta V(x) = V(x)-V(x^*) \geq 0
\]
We define the \textit{inverse effective temperature} as 
\begin{equation}\label{effb}
   \beta_{\text{eff}}(x)=\frac{1}{\cal E(x)}\log\frac{\rho(x^*)}{\rho(x)},\quad x\neq x^*
\end{equation}
where $\cal E(x)= E(x)-E(x^*)$.  It may be that there are multiple minima of $V$ and they may give rise to different effective temperatures; we just fix one choice.  In equilibrium, with \eqref{bg}, $\beta_{\text{eff}}(x) = \beta$.\\  
Writing in terms of the inverse bath temperature $\beta$, we have
\begin{eqnarray}
    C(\beta) &=& -\beta^2[ V(x^*)\frac{\partial}{\partial \beta} \rho (x^*)+\sum_{x\not=x^*}V(x)\frac{\partial}{\partial \beta} \rho(x)]\\
    &=& -\beta^2 \sum_{x\not=x^*}\Delta V(x)\frac{\partial}{\partial \beta} \rho(x)\\
  &=& -\beta^2 \sum_{x\not=x^*}\Delta V(x)\frac{\partial}{\partial\beta}[\rho(x^*)e^{-\cal E(x) \beta_{\text{eff}}(x)}] \notag\\
      &=&-\beta^2 \sum_{x\not=x^*}\Delta V(x)\,\rho(x^*)\,e^{-\cal E(x) \beta_{\text{eff}}(x)}[\frac{\partial\log\rho(x^*)}{\partial \beta} -\cal E(x) \frac{\partial \beta_{\text{eff}}(x)}{\partial\beta}]\\
      &=&\sum_{x\not=x^*}\Delta V(x)\,e^{-\cal E(x) \beta_{\text{eff}}(x)}\,[\frac{\partial\rho(x^*)}{\partial T} +\frac{\cal E(x) \,\rho(x^*)}{T_\text{eff}^2(x)}\,\frac{\partial T_{\text{eff}}(x)}{\partial T}]
    \end{eqnarray}
where we have used the normalization of $\rho$ in the second equality and Definition \eqref{effb} in the third equality.  The last two lines are just rewritings.\\
The situation is then as follows:\\  
When there exists a temperature range where $ \frac {\partial\rho(x^*)}{\partial T}$ is positive (unlike equilibrium), then a negative heat capacity will arise (and can only arise) from a sufficiently decreasing effective temperature $T_{\text{eff}}(x)$ in temperature $T$ (for some state $x$).  We illustrate that in the next section.  Indeed, in nonequilibrium, due to the driving,  population inversion can occur, and $ \frac {\partial\rho(x^*)}{\partial T}$ is easily made positive for some temperature range. \\
We can also get a negative heat capacity when $\frac{\partial\rho(x^*)}{\partial T} $ is negative (like in equilibrium), as long as  the slope of $T_\text{eff}$ against $T$ is sufficiently small: we  need that
$ \frac{\partial T_{\text{eff}}(x)}{\partial T} \leq  \frac{T^2_{\text{eff}}}{\rho(x^*)\, \cal E(x)}\,\left|\frac{\partial\rho(x^*)}{\partial T}\right|$ (for some state $x$). We will encounter that for instance in Figs.~\ref{hceff}--\ref{efft2} around $T=0.5$.  It is the situation when the effective temperature (almost) stops to depend on the environment temperature.\\

As a sufficient condition for negative heat capacity, we then have
\begin{equation}\label{suff}
\text{if} \,\,\,\forall x\not=x^*  \qquad  \cal E(x) \frac{\partial \beta_{\text{eff}}(x)}{\partial\beta}\leq \frac{\partial}{\partial \beta} \log \rho(x^*)  \,\quad  \Rightarrow C(\beta) \leq 0
\end{equation}
It is easy to verify that this never holds in equilibrium (using \eqref{bg}), but it is perfectly possible in a driven system.\\
We can rewrite \eqref{suff} using the heat bath temperature directly, for easier intuition,
\begin{align}\label{teffhc}
\text{if} \,\,\,\forall x\not=x^*  \qquad \frac{\partial\rho(x^*)}{\partial T} +\rho(x^*)\,\frac{\cal E(x)}{T^2_{\text{eff}}}\, \frac{\partial T_{\text{eff}}(x)}{\partial T}\leq 0 \quad  \Rightarrow C(T) \leq 0
\end{align}
We repeat that the effective temperature \eqref{effb} is defined with respect to {\it some} minimizer $x^*$ of the quasipotential, and we note that $x^* = x^*_T$ can also change with temperature $T$.   In many cases at low temperatures $x^* =  g$ where $g$ denotes the state that maximizes the pseudopotential $\Phi$ of \eqref{hcal}
and hence the stationary probability, $\rho_T(g)\geq \rho_T(x)$: the dominant state $g$ can then be identified with the state $x^*$, that minimizes the quasipotential.  That is certainly the case close-to-equilibrium, \cite{McL,closeheat}, but is not always true in nonequilibrium. 

\section{Illustrations}\label{illu}
We collect  a number of discrete examples to clearly illustrate the origin of negative heat capacity as discussed above.  We consider basically two regimes for the negativety, at very low, and at intermediate temperatures. The models are not suitable to study high-temperature physics, where we would need to turn to underdamped Langevin processes, possibly for an ideal gas with driven internal degrees of freedom.

\begin{example}[Example \ref{ex1} continued: negative heat capacity around zero temperature]\label{cont}
The low-temperature behavior of the heat capacity is
\[
C(T)  \simeq\frac{3}{2}\beta ^2 \varepsilon \alpha  (\varepsilon -2 \Delta )\, e^{ -\beta  \varepsilon}, \qquad T\downarrow 0
\]
which is negative for $\Delta>\ve/2$ (and $\alpha > 0$, which is the nonequilibrium case).  We see from Fig.~\ref{hceff} that $C(T)\leq 0$ for $T\leq 0.5$ when $\alpha=0.1, \ve=1, \Delta=1$. 
The analytic expressions are collected in Appendix \ref{3level}.\\
Observe that the quasipotential $V_T((1,-)  = x^*)$ is minimal at all temperatures $T$; see Fig.\ref{hceff}.  That minimizer for the quasipotential maximizes the pseudopotential, in the sense that $\rho(1,-)=\rho(3,+) \geq
 \rho(2,-)=\rho(2,+)\geq 
\rho(3,-)=\rho(1,+)$ for all temperatures as well.
\begin{figure}[H]
     \centering
      \begin{subfigure}{0.49\textwidth}
         \centering
         \def\svgwidth{0.8\linewidth}        
        \includegraphics[scale = 0.85]{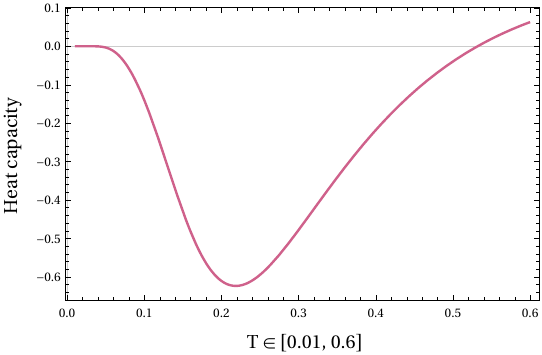}
     \end{subfigure}
     \hfill
     \begin{subfigure}{0.49\textwidth}
         \centering
         \def\svgwidth{0.8\linewidth}        
   \includegraphics[scale = 0.85]{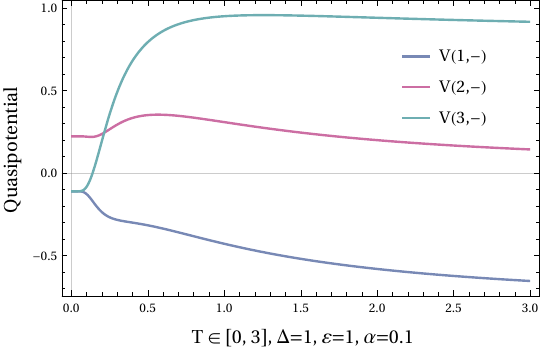}
     \end{subfigure}
     \caption{\small{Left: Heat capacity of  Example \ref{ex1} as a function of temperature at $\alpha=0.1, \ve=1, \Delta=1$. Right: $V_T(x)$ for the states $x$ of Example \ref{ex1}. The quasipotential $V_T(1,-)  = x^*$ is lowest at all temperatures. }}
\label{hceff}
\end{figure}
\begin{figure}
     \centering
      \begin{subfigure}{0.49\textwidth}
         \centering
         \def\svgwidth{0.8\linewidth}        
        \includegraphics[scale = 0.85]{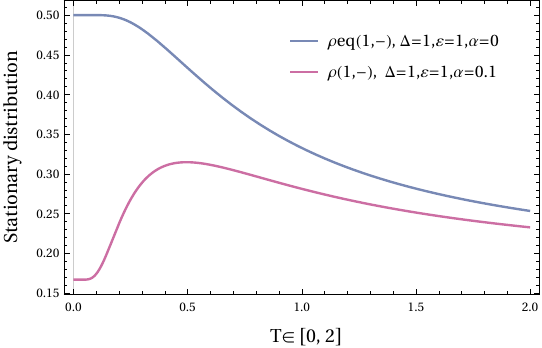}
     \end{subfigure}
     \hfill
     \begin{subfigure}{0.49\textwidth}
         \centering
         \def\svgwidth{0.8\linewidth}        
   \includegraphics[scale = 0.85]{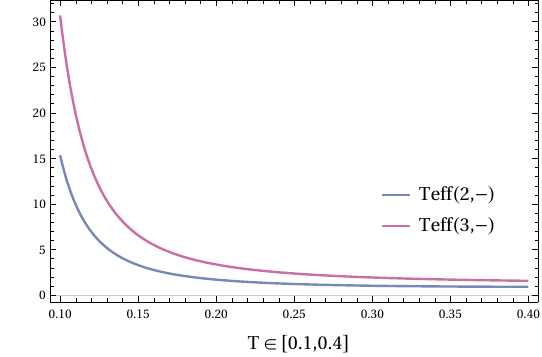}
     \end{subfigure}
     \caption{\small{Left: Stationary occupation of $(1,-)$ in Example \ref{ex1}, for varying temperatures. The top curve shows the equilibrium case ($\alpha=0$), and the lower curve is for  $\alpha=0.1$. In nonequilibrium there exists a range of temperatures where for $\Delta=1, \alpha=0.1$ and $\ve=1$ in Example \ref{ex1}, $\frac {\partial\rho(x^*)}{\partial T}>0$. Right: Effective temperatures  $T_{ \text{eff}}(2,-)$ and  $T_{ \text{eff}}(3,-)$ (defined from \eqref{effb}) are decreasing in the temperature range where the heat capacity is negative.}}
\label{efft2}
\end{figure}
Next, we see from \fig\ref{efft2}, that the stationary occupation of $x^* = (1,-)$ is increasing with temperature over $T < 0.5$, $ \frac {\partial\rho(1,-)}{\partial T} >0$, which is different from the equilibrium situation where the `ground state' loses occupation at higher temperature. At the same time,
as shown in the right figure, for the same temperature range,  $\frac{\partial T_{\text{eff}}(2,-)}{\partial T}<0$ and $\frac{\partial T_{\text{eff}}(3,-)}{\partial T}<0$. That produces the negative heat capacity according to \eqref{teffhc}.
\end{example}

\newpage
\begin{example}[uneven ladder: negative heat capacity at intermediate temperatures]\label{lad}
Consider the uneven 3-level ladder  in  \fig \ref{ladderstrange}, where the energies $E(\eta,\sigma)$ are
 \begin{align}\label{rateex2}
     E(1,-)=0, \qquad E(2,-)=2\ve, \qquad E(3,-)=3\ve.
 \end{align}
 and the transition rates are modified with an energy barrier $\Delta>0$,
 \begin{align*}
    & k_-(1,2)=k_+(3,2)=e^{-\beta\,( \Delta+\ve)}, \qquad  k_-(2,3)=k_+(2,1)=e^{-\beta\,( \Delta+\ve/2)}\\
      & k_+(1,2)=k_-(3,2)=e^{-\beta\,( \Delta+\ve)}, \qquad  k_+(2,3)=k_-(2,1)=e^{-\beta\,( \Delta+\ve/2)}\\
      &k_\eta(-,+)=k_\eta(+,-) =\alpha>0, \qquad \forall \eta=1,2,3.
 \end{align*}
  \begin{figure}[H]
         \centering
         \def\svgwidth{0.8\linewidth}        
        \includegraphics[scale = 0.7]{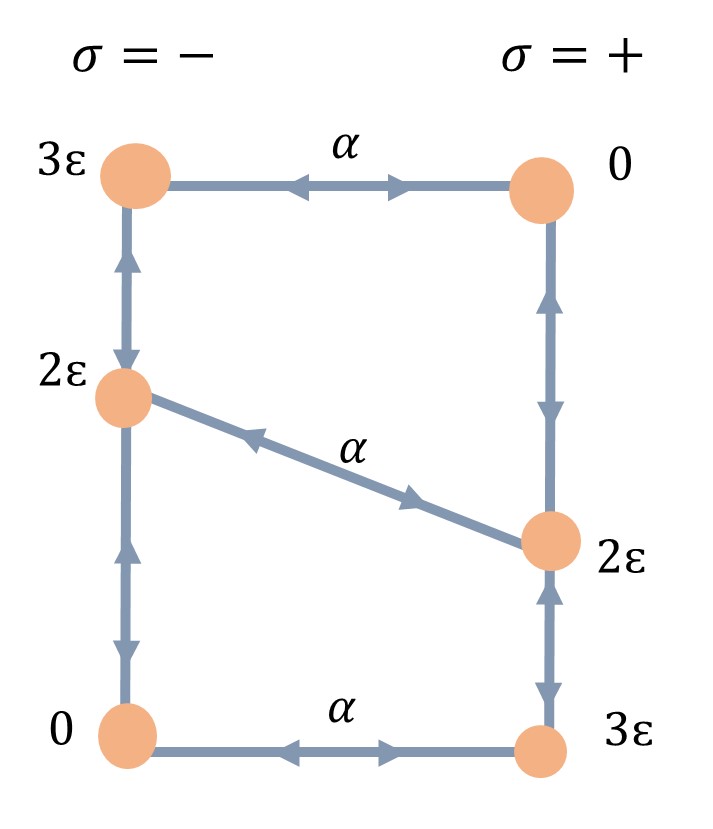}
\caption{\small Uneven {3-level ladder of Example \ref{lad}. }}   \label{ladderstrange}
\end{figure}
 The heat capacity of this system can become negative over intermediate values of temperature for certain values of $\alpha, \ve$ and $\Delta$; see \fig\ref{hcnonsy}, where the heat capacity is plotted for fixed $\alpha=0.1, \ve=2$ and different $\Delta$. 
\begin{figure}[H]
     \centering
      \begin{subfigure}{0.49\textwidth}
         \centering
         \def\svgwidth{0.8\linewidth}        
        \includegraphics[scale = 0.7]{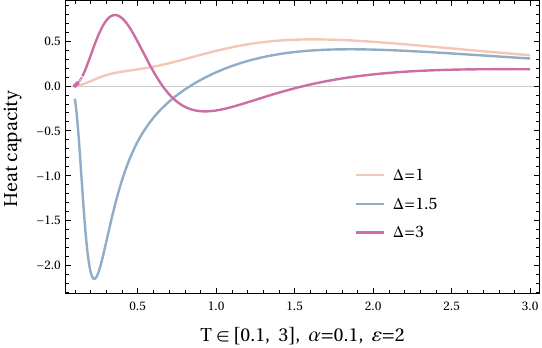}
     \end{subfigure}
     \hfill
     \begin{subfigure}{0.49\textwidth}
         \centering
         \def\svgwidth{0.8\linewidth}        
   \includegraphics[scale = 0.7]{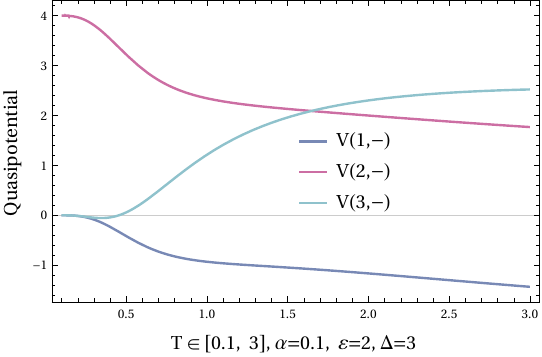}
     \end{subfigure}
     \caption{\small{Left: Heat capacity of Example \ref{ex2} for varying temperature, different values of $\Delta$ and fixed values of   $\alpha=0.1,\, \ve=2$. Right: Quasipotentials at $\Delta=3$,  $ \alpha=0.1$and $\ve=2$ , where $x^*=(1,-)$. }}
\label{hcnonsy}
\end{figure}
In \fig \ref{hcnonsy}, the heat capacity remains positive for $\Delta=1$. For higher values of $\Delta$, it becomes negative at low temperatures and for even higher values of $\Delta$, it becomes negative at higher (intermediate) temperatures.\\
For an understanding in terms of \eqref{teffhc}, we observe from \fig \ref{rhodelta} that the stationary occupation of $(1,-)$ can be increasing with temperature for high enough $\Delta$. On the other hand, the effective temperature of the other states located on the leg $-$ are seen for $\Delta=3$ in the right plot. For the intermediate range of temperatures where the heat capacity is negative,  $T_{\text{eff}}(3,-)$ is strongly decreasing in temperature. 

\begin{figure}[H]
     \centering
      \begin{subfigure}{0.49\textwidth}
         \centering
         \def\svgwidth{0.8\linewidth}        
        \includegraphics[scale = 0.8]{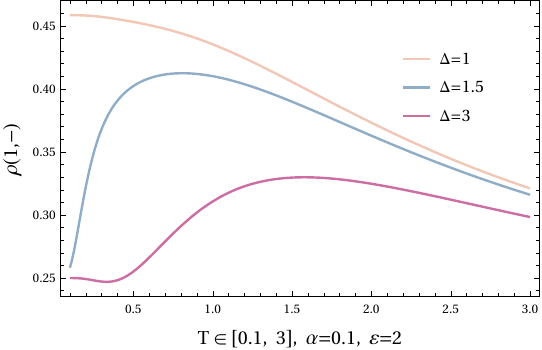}
     \end{subfigure}
     \hfill
     \begin{subfigure}{0.49\textwidth}
         \centering
         \def\svgwidth{0.8\linewidth}        
   \includegraphics[scale = 0.8]{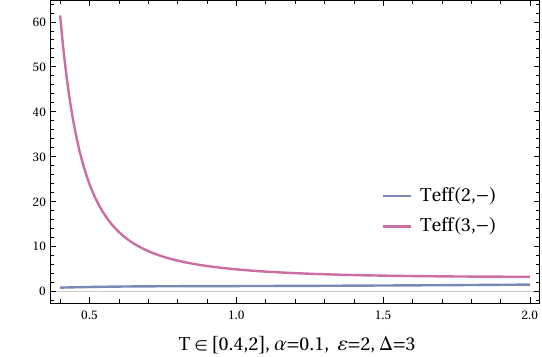}
     \end{subfigure}
     \caption{\small{Left: stationary distribution of state $(1,-)$ for different values of $\Delta$ for the Example \ref{lad} with $\alpha=0.1, \ve=2$,   $\rho(x^*)$ can be increasing with temperature  for higher values of $\Delta$. Right: the effective temperatures with $T_{\text{eff}}(3,-)$ strongly decreasing in the intermediate range of temperatures where the heat capacity is negative.  }}
\label{rhodelta}
\end{figure}
\end{example}

\begin{example}[double-channel two-level model with negative heat capacity at intermediate temperatures]\label{ex2}\label{karel}
    Consider the two-level system with states $1$ and $2$, with possible transitions over two channels, $+$ and $-$. See \fig~\ref{twochann}, and \cite{cejp}.
    \begin{figure}[H]
         \centering
         \def\svgwidth{0.8\linewidth}        
        \includegraphics[scale = 0.7]{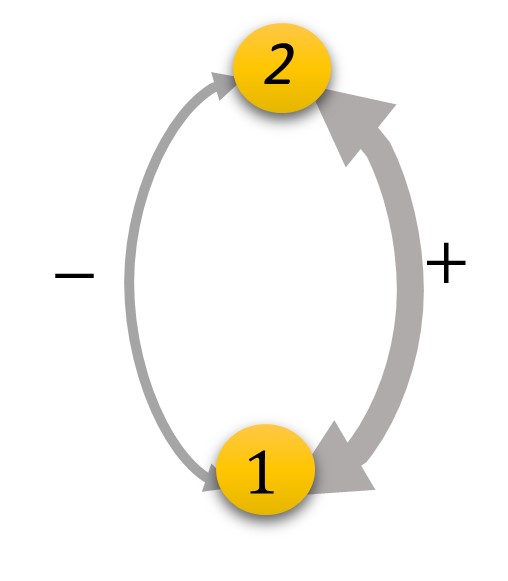}
\caption{\small{Double-channel 2level-model of Example \ref{karel}}.  The plus-channel is more reactive than the minus-channel for $\phi>0$ in \eqref{2r}.}   \label{twochann}
\end{figure}
The transition rates are 
    \begin{align}\label{2r}
        k_-(1,2)&=e^{- \beta(w+\ve)/2},\qquad k_-(2,1)=e^{\beta(w+\ve)/2} \notag\\
        k_+ (1,2)&=e^{\phi+ \beta(w-\ve)/2},\qquad k_+(2,1)=e^{\phi-\beta(w-\ve)/2}
    \end{align}
   where for instance $k_+(1,2)$ denotes the transition rate from state $1$ to state $2$ through the channel $+$. The nonequilibrium driving is $w$, and the energy difference is given by $\cal E(2)=\ve$.  In equilibrium ($w=0$), $\rho(1)=\frac{e^{\beta  \ve}}{e^{\beta  \ve}+1}$ and $\rho(2)=\frac{1}{e^{\beta  \ve}+1}$.  The analytic expressions for the nonequilibrium stationary distribution, quasipotential and dissipated heat are provided in Appendix \ref{2c}. \\
In \fig \ref{twochannelhc} and \fig \ref{hceff4chan}, the heat capacity is plotted for different values of $\phi$ and $w$ in varying temperatures. As it is shown in higher values of driving $w$ the heat capacity can be negative at intermediate temperatures; see\fig \ref{twochannelhc}.  The mechanism is different from the previous example (where the barrier was the relevant variable for the nonequilibrium regime); here it is the asymmetry between the two channels that does the job (of negative heat capacity).
\begin{figure}[H]
     \centering
      \begin{subfigure}{0.49\textwidth}
         \centering
         \def\svgwidth{0.8\linewidth}        
        \includegraphics[scale = 0.85]{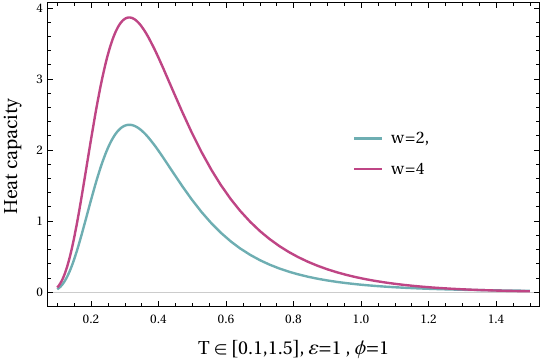}
     \end{subfigure}
     \hfill
     \begin{subfigure}{0.49\textwidth}
         \centering
         \def\svgwidth{0.8\linewidth}        
   \includegraphics[scale = 0.85]{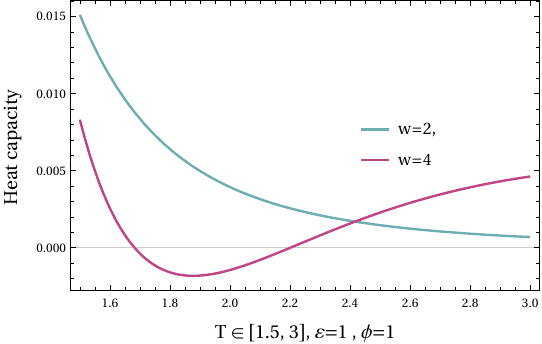}
     \end{subfigure}
\caption{\small{ The heat capacity of Example \ref{karel} for  $\ve=\phi=1$ at different values of $w$. Left: the heat capacity at low temperature remains positive for both $w=2$ and $w=4$. Right: the heat capacity at a higher temperature is positive for $w=2$  and is negative for $w=4$.}  } \label{twochannelhc}
\end{figure}
The analytical expression of the heat capacity is given in Appendix \ref{2c}, and  there is a transition in $\phi$, depending on the driving $W$, between completely positive and partally negative temperature-regimes of the heat capacity.
\begin{figure}[H]
     \centering
      \begin{subfigure}{0.49\textwidth}
         \centering
         \def\svgwidth{0.8\linewidth}        
        \includegraphics[scale = 0.85]{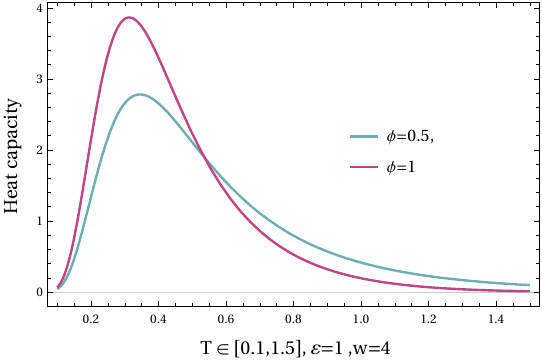}
     \end{subfigure}
     \hfill
     \begin{subfigure}{0.49\textwidth}
         \centering
         \def\svgwidth{0.8\linewidth}        
   \includegraphics[scale = 0.85]{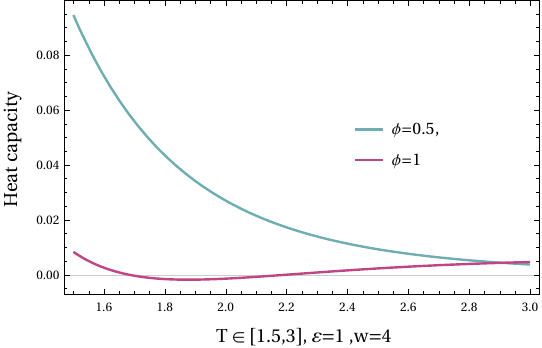}
     \end{subfigure}
\caption{\small{The heat capacity of Example \ref{karel} for  $\ve=1, w=4$ at different values of $\phi$. Left: the heat capacity at low temperature remains positive for both $\phi=0.5$ and $\phi=1$. Right: the heat capacity at a higher temperature is positive for $\phi=0.5$  and is negative for $\phi=1$. }}   \label{hceff4chan}
\end{figure}
The quasipotential is plotted in \fig \ref{quasi2channel} as a function of temperature.
\begin{figure}[H]
         \centering
         \def\svgwidth{0.8\linewidth}        
        \includegraphics[scale = 0.9]{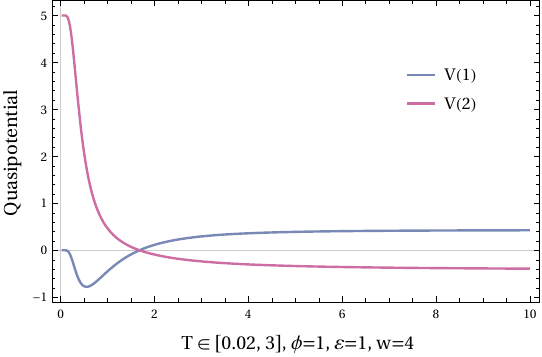}
\caption{\small{Quasipotential for Example \ref{karel} at $w=4$ and $\phi=\ve=1$. }}   \label{quasi2channel}
\end{figure}
$V(2) < V(1)$ for $w=4$, $\phi=\ve=1$ and $T\in [1.7,2.2]$, and  $x^*=2$. For the negativity of the heat capacity we see from \fig \ref{rho2chan} that $\frac{\partial \rho(2)}{\partial T}>0$ is positive for $T\in [1.7,2.2]$.  We have plotted the effective temperature of state $2$ in \fig \ref{Tefftwolevel}.
\begin{figure}[H]
   \includegraphics[scale = 0.43]{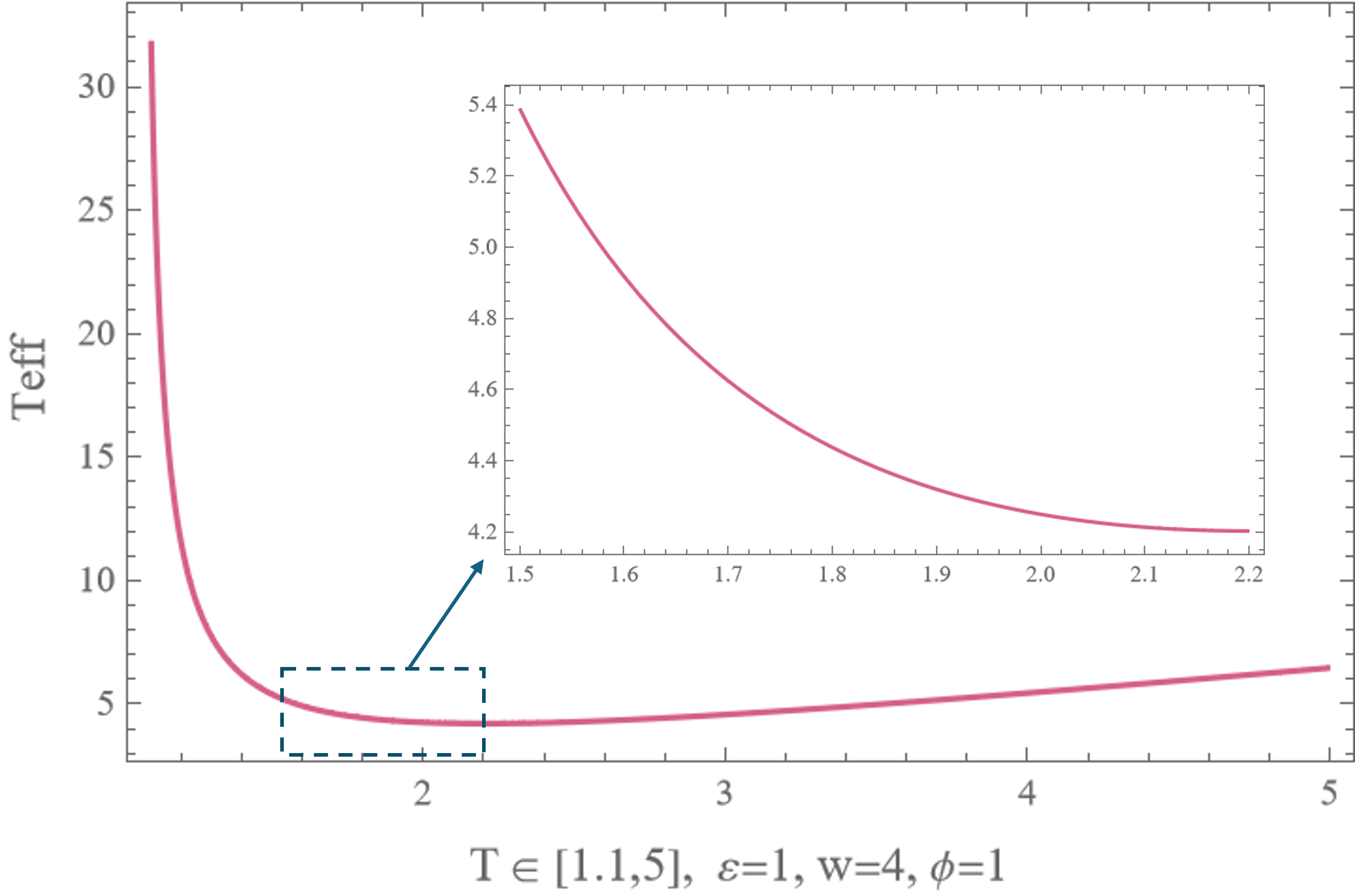}
\caption{\small{The effective temperature of state $1$ with respect to temperature for Example \ref{karel}. The right plot is the zoom in the range of temperatures where the heat capacity is negative. }}   \label{Tefftwolevel}
\end{figure}
Again, where the effective temperature is decreasing with temperature we get a negative heat capacity; see \fig \ref{twochannelhc}.
\begin{figure}[H]
     \centering      
  \includegraphics[scale = 0.4]{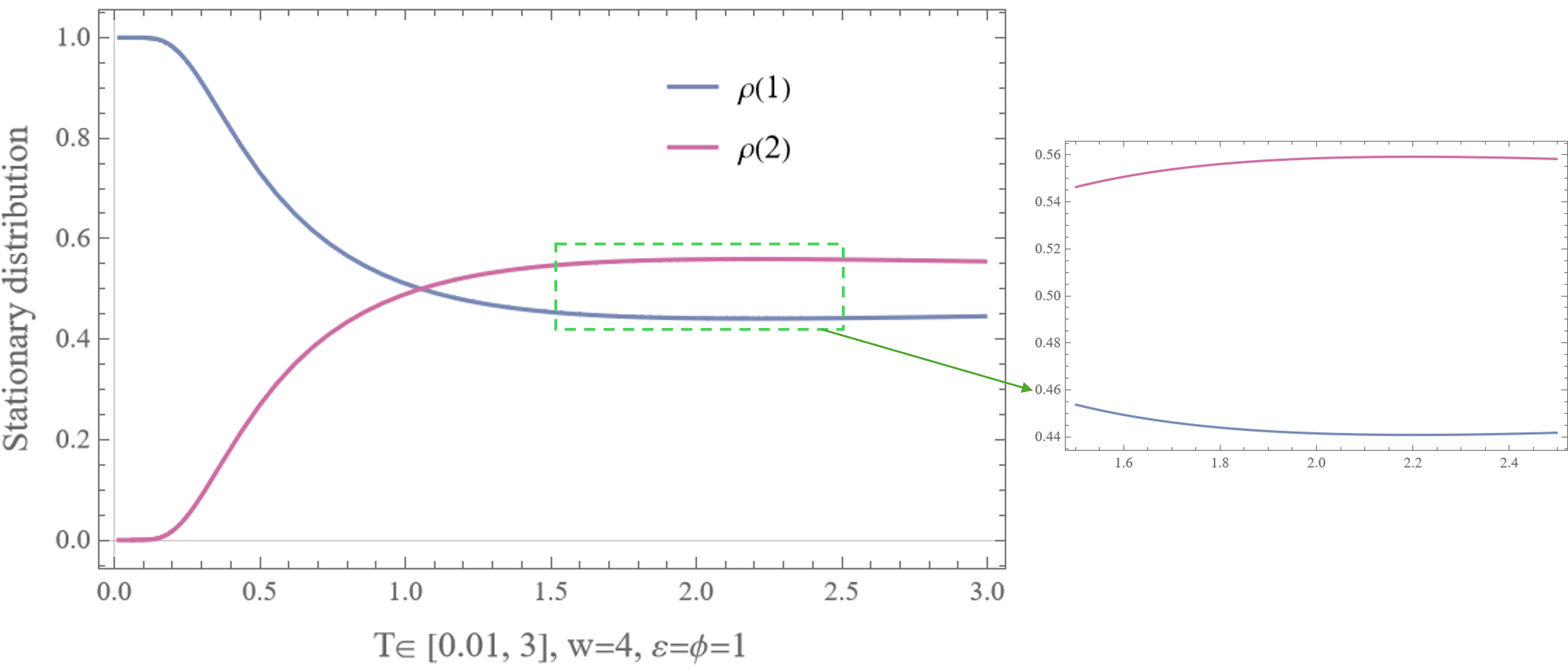}
\caption{\small{Stationary distribution for Example \ref{karel}, with $T\in [2, 10]$ at $w=4$ and $\phi=\ve=1$. It is important that $\frac{\partial \rho(1)}{\partial T}>0$. }}   \label{rho2chan}
\end{figure}

Observe from \fig \ref{quasi2channel}  that although $x^*=1$, the stationary probability $\rho(1)$ is not the highest at all temperatures.
\end{example}

\section{Conclusion}
For steady nonequilibrium systems, the specific heat at fixed volume is not immediately related to the change of the total average energy with respect to changes in kinetic energy.  The latter can be negative for certain equilibrium systems.  Instead, the negativity of the specific heat for driven or active systems arises from an anticorrelation between the expected absorbed heat for a given system condition and the change in its population level for increased bath temperature. The former relates to the Clausius entropy, and the latter connects to the Boltzmann entropy.  Those entropies do not march together when out of equilibrium, and that is the ultimate reason for the negativity of nonequilibrium heat capacities.  The present paper substantiates that claim in quantitative ways and various examples have been used to illustrate the possible scenarios.
\vspace{1cm}

\noindent {\bf Acknowledgment:}
We are grateful to Karel Neto\v{c}n\'{y} for many discussions that made this work possible.

\newpage
\appendix
\section{Three-level ladder}\label{3level}
We include  the explicit formulae for Example \ref{ex1}.\\

\begin{figure}
     \centering
      \begin{subfigure}{0.49\textwidth}
         \centering
         \def\svgwidth{0.8\linewidth}        
        \includegraphics[scale = 0.85]{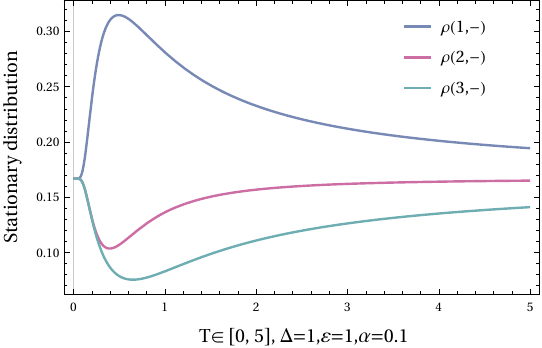}
     \end{subfigure}
     \hfill
     \begin{subfigure}{0.49\textwidth}
         \centering
         \def\svgwidth{0.8\linewidth}        
   \includegraphics[scale = 0.85]{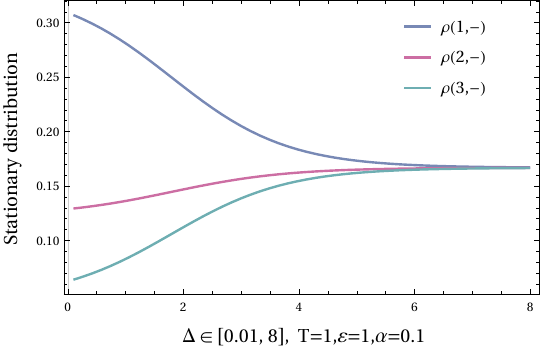}
     \end{subfigure}
      \centering
      \begin{subfigure}{0.49\textwidth}
         \centering
         \def\svgwidth{0.8\linewidth}        
        \includegraphics[scale = 0.85]{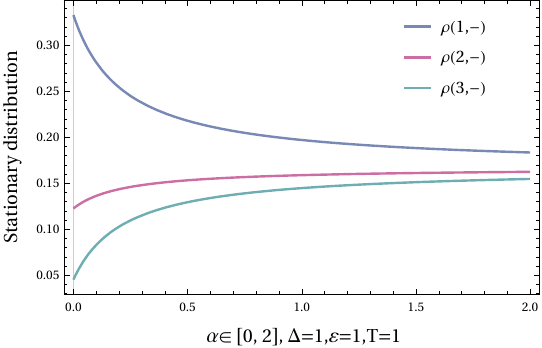}
     \end{subfigure}
\caption{\small{Stationary distributions of the three-level ladder given in  Example \ref{ex1} for different values of $\alpha, \Delta $ and temperature $T$. }}   \label{rhoplots}
\end{figure}

Solving the stationary  Master Equation for the rates given in \eqref{rate1}, the stationary distribution is 
\begin{align*}
\rho(1,-)=\rho(3,+)&=\frac{1}{Z}e^{\frac{\beta  \varepsilon }{2}}  \left(\alpha  e^{\beta  (\Delta +\varepsilon )}+\alpha  e^{\beta  \Delta }+e^{\frac{3 \beta  \varepsilon }{2}}\right)\geq\\
 \rho(2,-)=\rho(2,+)&=\frac{1}{Z}\left(\alpha  e^{\beta  \left(\Delta +\frac{3 \varepsilon }{2}\right)}+\alpha  e^{\beta  \Delta +\frac{\beta  \varepsilon }{2}}+e^{\beta  \varepsilon }\right) \geq\\ 
\rho(3,-)=\rho(1,+)&=\frac{1}{Z}\left(2 \alpha  e^{\beta  (\Delta +\varepsilon )} \cosh \left(\frac{\beta  \varepsilon }{2}\right)+1\right),
\end{align*}
where 
\[Z=2 e^{\frac{\beta  \varepsilon }{2}} \left(e^{\beta  \varepsilon }+1\right) \left(3 \alpha  e^{\beta  \Delta }+e^{\frac{\beta  \varepsilon }{2}}\right)+2.\]
We plot the stationary distribution of the states on the leg $\sigma=-$ for different values of $\alpha, \Delta$ and temperature; see \fig \ref{rhoplots}. The ground states $(1,-)$ and $(3,+)$ are the most dominant states. Toward zero temperature for values of $\Delta >\ve/2$ all the transition rates become equal (approaching zero); consequently, all states become equivalent.\\

The quasipotential equals 
\begin{align*}
V(1,-)=V(3,+)&=-\frac{1}{N}\varepsilon  \bigg[\alpha ^2 \left(-e^{\beta  (2 \Delta +\varepsilon )}\right)+\alpha ^2 e^{2 \beta  \Delta +3 \beta  \varepsilon }+5 \alpha  e^{\beta  \left(\Delta +\frac{3 \varepsilon }{2}\right)}+7 \alpha  e^{\beta  \left(\Delta +\frac{5 \varepsilon }{2}\right)}\\
&\quad+3 \alpha  e^{\beta  \left(\Delta +\frac{7 \varepsilon }{2}\right)}+3 \alpha  e^{\beta  \Delta +\frac{\beta  \varepsilon }{2}}+3 e^{\beta  \varepsilon }+3 e^{2 \beta  \varepsilon }+e^{3 \beta  \varepsilon }+2\bigg],
\end{align*}
\begin{align*}
V(2,-)=V(2,+)&=\frac{1}{N}\varepsilon  \left(e^{\beta  \varepsilon }-1\right) \bigg[2 \alpha ^2 e^{2 \beta  (\Delta +\varepsilon )}+2 \alpha ^2 e^{\beta  (2 \Delta +\varepsilon )}+\alpha  e^{\beta  \left(\Delta +\frac{5 \varepsilon }{2}\right)}\\
&\quad +\alpha  e^{\beta  \Delta +\frac{\beta  \varepsilon }{2}}+2 e^{\beta  \varepsilon }+2 e^{2 \beta  \varepsilon }+e^{3 \beta  \varepsilon }+1\bigg],\\
V(3,-)=V(1,+)&=\frac{1}{N}\varepsilon  e^{\frac{\beta  \varepsilon }{2}} \bigg[\alpha ^2 \left(-e^{2 \beta  \Delta +\frac{5 \beta  \varepsilon }{2}}\right)+\alpha ^2 e^{\frac{1}{2} \beta  (4 \Delta +\varepsilon )}+7 \alpha  e^{\beta  (\Delta +\varepsilon )}+5 \alpha  e^{\beta  (\Delta +2 \varepsilon )}\\
&+3 \alpha  e^{\beta  (\Delta +3 \varepsilon )}+3 \alpha  e^{\beta  \Delta }+e^{\frac{\beta  \varepsilon }{2}}+3 e^{\frac{3 \beta  \varepsilon }{2}}+3 e^{\frac{5 \beta  \varepsilon }{2}}+2 e^{\frac{7 \beta  \varepsilon }{2}}\bigg],
\end{align*}
where 
\[N=2\, [e^{\frac{\beta  \varepsilon }{2}} (e^{\beta  \varepsilon }+1) (3 \alpha  e^{\beta  \Delta }+e^{\frac{\beta  \varepsilon }{2}})+1]^2.\]
Since the steady average of the quasipotentials is zero, there must be at least one state with a negative quasipotential. In this example, the states with the highest stationary occupation correspond to those negative quasipotentials. \\

The heat capacity is computed from \eqref{ct} to be
\begin{align*}
    C(\beta)=&\frac{\beta ^2 \varepsilon  e^{\frac{\beta  \varepsilon }{2}}}{2 \left(e^{\frac{\beta  \varepsilon }{2}} \left(e^{\beta  \varepsilon }+1\right) \left(3 \alpha  e^{\beta  \Delta }+e^{\frac{\beta  \varepsilon }{2}}\right)+1\right)^3} \Bigg[-4 \alpha ^2 (\Delta -\varepsilon ) e^{2 \beta  \Delta +\frac{3 \beta  \varepsilon }{2}}-\alpha ^2 (2 \Delta -\varepsilon ) e^{2 \beta  \Delta +\frac{9 \beta  \varepsilon }{2}}\\
    &-10 \alpha ^2 \varepsilon  e^{2 \beta  \Delta +\frac{5 \beta  \varepsilon }{2}}+4 \alpha ^2 (\Delta +\varepsilon ) e^{2 \beta  \Delta +\frac{7 \beta  \varepsilon }{2}}+\alpha ^2 (2 \Delta +\varepsilon ) e^{\frac{1}{2} \beta  (4 \Delta +\varepsilon )}\\
    &+3 \alpha  (\varepsilon -2 \Delta ) e^{\beta  (\Delta +5 \varepsilon )}+3 \alpha  e^{\beta  \Delta } (2 \Delta +\varepsilon )+2 \alpha  (6 \Delta +7 \varepsilon ) e^{\beta  (\Delta +\varepsilon )}+\alpha  (19 \varepsilon -6 \Delta ) e^{\beta  (\Delta +3 \varepsilon )}\\
    &+\alpha  (6 \Delta +19 \varepsilon ) e^{\beta  (\Delta +2 \varepsilon )}-2 \alpha  (6 \Delta -7 \varepsilon ) e^{\beta  (\Delta +4 \varepsilon )}\\
    &+2 \varepsilon  e^{\frac{\beta  \varepsilon }{2}}+10 \varepsilon  e^{\frac{3 \beta  \varepsilon }{2}}+12 \varepsilon  e^{\frac{5 \beta  \varepsilon }{2}}+10 \varepsilon  e^{\frac{7 \beta  \varepsilon }{2}}+2 \varepsilon  e^{\frac{9 \beta  \varepsilon }{2}} \Bigg]
\end{align*}
At very low temperatures with $\Delta>0$,
\begin{align*}
   \lim_{\beta\to \infty} C(\beta)\sim&\frac{\beta ^2 \varepsilon  e^{\frac{\beta  \varepsilon }{2}}}{2 \left(3 e^{\frac{3\beta  \varepsilon }{2}+\beta \Delta} +e^{\frac{4 \beta  \varepsilon }{2}}\right)^3} \big(-\alpha^2(2\Delta -\ve)e^{2 \beta \Delta +\frac{9 \beta \ve}{2}} +3 \alpha  (\varepsilon -2 \Delta ) e^{\beta  (\Delta +5 \varepsilon )}\big)\\
   \sim&\frac{3\beta ^2 \varepsilon \alpha  (\varepsilon -2 \Delta ) }{2 e^{ \beta  \varepsilon} }, 
   \end{align*}
With $\alpha \ve>0$, at low temperatures, the heat capacity is negative for $\Delta>\ve/2$.
\section{Two-channel model}\label{2c}
We collect the formulae for Example~\ref{karel}; see also \cite{cejp}.\\

The stationary distribution is 
\begin{align}
    \rho(1)=\frac{e^{\beta w }+e^{\phi }}{e^{\beta w }+e^{\beta w +\beta  (-\ve)+\phi }+e^{\beta  (-\ve)}+e^{\phi }}, \qquad \rho(2)=\frac{e^{\beta w +\phi }+1}{e^{\beta w +\phi }+e^{\beta  (w +\ve)}+e^{\beta  \ve+\phi }+1}
\end{align}
The quasipotential is obtained from solving the Poisson equation \eqref{poisson},
\begin{align*}
    V(1)&=-\frac{\left(e^{\beta  w +\phi }+1\right) \left(e^{\phi } (\ve-w ) \left(e^{\beta  w }+e^{\beta  \ve}\right)+(w +\ve) \left(e^{\beta  (w +\ve)}+1\right)\right)}{\left(e^{\beta  w +\phi }+e^{\beta  (w +\ve)}+e^{\beta  \ve+\phi }+1\right)^2} \\
   V(2)&=\frac{e^{\beta  \ve} \left(e^{\beta  w }+e^{\phi }\right) \left(e^{\phi } (\ve-w ) \left(e^{\beta  w }+e^{\beta  \ve}\right)+(w +\ve) \left(e^{\beta  (w +\ve)}+1\right)\right)}{\left(e^{\beta  w +\phi }+e^{\beta  (w +\ve)}+e^{\beta  \ve+\phi }+1\right)^2},
\end{align*}
where the expected dissipated heat fluxes equal
\begin{align*}
\cal P(1)&=(w -\ve) \exp [\frac{1}{2} \beta  (w -\ve)+\phi ]+(-w -\ve) \exp [\frac{1}{2} \beta  (-w -\ve)]\\
\cal P(2)&=-(w -\ve) \exp [\phi -\frac{1}{2} \beta  (w -\ve)]-(-w -\ve) \exp [\frac{1}{2} (-\beta ) (-w -\ve)].
\end{align*}
Hence, the heat capacity is
\begin{align*}
C(\beta)=&\frac{1}{a}2 \beta ^2 e^{\beta  (w +\ve)+\phi } \left(e^{\phi } (\ve-w ) \left(e^{\beta  w }+e^{\beta  \ve}\right)+(w +\ve) \left(e^{\beta  (w +\ve)}+1\right)\right) \\
&\times(\ve (\cosh (\beta  w )+\cosh (\phi ))-w  \sinh (\phi ))\\
a=&\left(e^{\beta  w +\phi }+e^{\beta  (w +\ve)}+e^{\beta  \ve+\phi }+1\right)^3
\end{align*}
For $w>\ve\geq0$ and all values of $\phi$ and $\beta$, if  $e^{\phi } (w-\ve ) \left(e^{\beta  w }+e^{\beta  \ve}\right)>(w +\ve) \left(e^{\beta  (w +\ve)}+1\right)$ and $\ve (\cosh (\beta  w )+\cosh (\phi ))>w  \sinh (\phi )$, then the heat  capacity is negative. 
Another condition  that leads to negative heat capacity is   $e^{\phi } (w-\ve ) \left(e^{\beta  w }+e^{\beta  \ve}\right)<(w +\ve) \left(e^{\beta  (w +\ve)}+1\right)$ and $\ve (\cosh (\beta  w )+\cosh (\phi ))<w  \sinh (\phi )$.  For $\phi=0$; the heat capacity is always positive for $w>\ve\geq0$.\\

At very low temperatures, 
\[C(\beta)\sim \frac{2 \beta ^2 (w +\ve)\,e^{2\beta  (w +\ve)+\phi }\, \ve e^{\beta w} }{e^{3\beta  (w +\ve)}}\sim 2 \ve \beta ^2 (w +\ve)\,e^{-\beta \ve+\phi }\]
     \bibliographystyle{unsrt}  
\bibliography{chr}
\onecolumngrid
\end{document}